\documentclass{emulateapj}
\usepackage{amsmath}
\usepackage[dvips]{color}

\def\stacksymbols #1#2#3#4{\def\theguybelow{#2}
        \def\verticalposition{\lower#3pt}
        \def\spacingwithinsymbol{\baselineskip0pt\lineskip#4pt}
        \mathrel{\mathpalette\intermediary#1}}
\def\intermediary #1#2{\verticalposition\vbox{\spacingwithinsymbol
        \everycr={}\tabskip0pt
        \halign{$\mathsurround0pt#1\hfil##\hfil$\crcr#2\crcr
                \theguybelow\crcr}}}

\shorttitle{FORMATION OF THE \emph{FERMI} BUBBLES IN A VISCOUS HALO}
\shortauthors{GUO ET AL.}

\begin{document}
\bibliographystyle{apj} 

\title {THE FERMI BUBBLES. II. THE POTENTIAL ROLES OF VISCOSITY AND COSMIC RAY DIFFUSION IN JET MODELS}

\author{Fulai Guo\altaffilmark{1}, William G. Mathews \altaffilmark{1}, Gregory Dobler\altaffilmark{2}, and S. Peng Oh\altaffilmark{3}}

\altaffiltext{1}{UCO/Lick Observatory, Department of Astronomy and Astrophysics, University of California, Santa Cruz, CA 95064, USA; fulai@ucolick.org}
\altaffiltext{2}{Kavli Institute for Theoretical Physics, University of California, Santa Barbara Kohn Hall, Santa Barbara, CA 93106, USA}
\altaffiltext{3}{Department of Physics, University of California, Santa Barbara Kohn Hall, Santa Barbara, CA 93106, USA}

\begin{abstract}

The origin of the \emph{Fermi} bubbles recently detected by the {\it Fermi Gamma-ray Space Telescope} in the inner Galaxy is mysterious. In the companion paper Guo \& Mathews (Paper I), we use hydrodynamic simulations to show that they could be produced by a recent powerful AGN jet event. Here we further explore this scenario to study the potential roles of shear viscosity and cosmic ray (CR) diffusion on the morphology and CR distribution of the bubbles. We show that even a relatively low level of viscosity ($\mu_{\rm visc} \gtrsim 3\text{  g cm}^{-1}\text{ s}^{-1}$, or $\sim 0.1\%$ - $1\%$ of Braginskii viscosity in this context) could effectively suppress the development of Kelvin-Helmholtz instabilities at the bubble surface, resulting in smooth bubble edges as observed. Furthermore, viscosity reduces circulating motions within the bubbles, which would otherwise mix the CR-carrying jet backflow near bubble edges with the bubble interior. Thus viscosity naturally produces an edge-favored CR distribution, an important ingredient to produce the observed flat gamma-ray surface brightness distribution. Generically, such a CR distribution often produces a limb-brightened gamma-ray intensity distribution. However, we show that by incorporating CR diffusion which is strongly suppressed across the bubble surface (as inferred from sharp bubble edges) but is close to canonical values in the bubble interior, we obtain a reasonably flat gamma-ray intensity profile. The similarity of the resulting CR bubble with the observed \emph{Fermi} bubbles strengthens our previous result in Paper I that the \emph{Fermi} bubbles were produced by a recent AGN jet event. Studies of the nearby \emph{Fermi} bubbles may provide a unique opportunity to study the potential roles of plasma viscosity and CR diffusion on the evolution of AGN jets and bubbles.
\end{abstract}


\section{Introduction}
\label{section:intro}

The recent discovery of two large gamma-ray bubbles by the \emph{Fermi Gamma-Ray Space Telescope} (\citealt{dobler10}; \citealt{su10}) has significantly changed the big picture of our Galaxy, the Milky Way. These ``\emph{Fermi} bubbles'' emit at $1\lesssim E_{\gamma}\lesssim 100$ GeV in the inner Galaxy, are nearly symmetric about the Galactic plane, and extend to $\sim 50^{\circ}$ ($\sim 10$ kpc) above and below the Galactic center (GC), with a width of about $40^{\circ}$ in longitude.  In addition, they have approximately uniform gamma-ray surface brightness with sharp edges.

The origin of the \emph{Fermi} bubbles has recently received a lot of attention and is actively debated in the literature. The sharp edges and bilobular morphology of the bubbles make them difficult to be explained by either diffused CRs from the Galactic disk or annihilations of dark matter particles (though \citealt{dobler11} claimed that dark matter annihilations in a prolate halo with anisotropic CR diffusion may explain the latter feature). \citet{crocker11} suggested that the gamma ray emission is powered by CR protons, which are continuously injected by supernova explosions in the Galactic center during a few Gyrs, though this seems difficult to reconcile with the sharp edges which implies a more recent transient event. In contrast, \citet{dobler10} and \citet{su10} argued that the emission may be dominated by upscattering of photons in the interstellar radiation field (ISRF) and the cosmic microwave background by CR electrons, whose synchrotron emission may have already been detected at tens of GHz by the {\it Wilkinson Microwave Anisotropy Probe} (WMAP; \citealt{finkbeiner04a}; \citealt{dobler08}). \citet{cheng11} argued that periodic star capture processes by the Galactic supermassive black hole, Sgr A$^{*}$, release AGN winds into the Galactic halo, producing the \emph{Fermi} bubbles within a few Myrs. \citet{zubovas11} suggested that the near-spherical outflow from a quasar event of Sgr A$^{*}$ around $6$ Myr ago may explain the origin of the bubbles.

In a companion paper (\citealt{guo12}; hereafter denoted as ``Paper I"), we performed the first numerical simulation following the dynamical evolution of the \emph{Fermi} bubbles, and showed that a recent AGN jet event originated from Sgr A$^{*}$ around $1$ - $3$ Myr ago can reproduce the \emph{Fermi} bubbles with roughly the observed location, size, and shape. Our jet model is inspired by many extragalactic AGN jets, which are clearly producing CR-filled bubbles seen in radio observations \citep{mcnamara07}. In our model, the opposing jets, dominated by kinetic energy and over-pressured by either CR or thermal pressure, were active for $\sim 0.1$ - $0.5$ Myr and moderately light. We also show that the sharp bubble edges require that CR diffusion across the bubble edges is suppressed significantly below the CR diffusion rate estimated in the solar vicinity.

While many observational features of the \emph{Fermi} bubbles, particularly their age, location, size and shape, are reproduced by our model in Paper I, the simulated bubble is deficient in two important respects -- surface irregularities and limb darkening in gamma ray surface brightness, disagreeing significantly with smooth edges and roughly uniform gamma ray surface brightness of the observed \emph{Fermi} bubbles. However, these inconsistencies do not necessarily mean that the jet scenario for the \emph{Fermi} bubbles is wrong. Instead, they may be smoking-gun signatures of additional physics, which plays a significant role during the jet evolution. Surface irregularities induced by Kelvin-Helmholtz instabilities have previously been seen in hydrodynamical simulations of buoyantly-rising X-ray cavities in galaxy clusters, where additional physical mechanisms, including hot gas viscosity \citep{reynolds05,kaiser05} and magnetic draping \citep{lyutikov06,ruszkowski07,dursi08}, have been invoked to suppress the instabilities. The flatness of the line-of-sight projected gamma ray intensity distribution is not trivial, as a spatially-uniform CR distribution produces a center-brightened gamma ray surface brightness while an edge-dominated CR distribution produces \emph{limb} brightening. The difficulties involved in producing such a `flat' gamma ray brightness in the \emph{Fermi} bubbles were previously noticed by \citet{mertsch11}, who suggested that it may be achieved if CR electrons are re-accelerated preferentially near bubble edges. 

In this paper, we further explore our jet model for the \emph{Fermi} bubbles developed in Paper I by including additional gas microphysics -- shear viscosity. We show that even a relatively low level of viscosity (compared to Spitzer viscosity in the shock-heated surrounding gas) significantly affects the evolution of the resulting \emph{Fermi} bubbles, helping produce smooth bubble edges and a flat gamma ray surface brightness as observed. We investigate the roles of viscosity and CR diffusion on the morphology and CR distribution of the bubbles by directly following the jet evolution, which differs significantly from \citet{reynolds05}, who investigated the role of viscosity on the buoyant rise of initially static bubbles in galaxy clusters. Furthermore, our paper is the first to study the role of viscosity in the context of the \emph{Fermi} bubbles.
 
The rest of the paper is organized as follows. In Section~\ref{section2}, we describe our model and numerical methods. We present our results in Section~\ref{section:results}, and summarize our main conclusions with implications in Section~\ref{section:conclusion}. In the Appendix, we explicitly present how we implement shear viscosity in cylindrical coordinates with axisymmetry.
  
\section{The Model and Numerical Methods}
\label{section2}

In Paper I and this paper, we study the formation of the \emph{Fermi} bubbles in the Milky Way's potential well with AGN jets using numerical simulations. We presented the basic picture of the jet scenario in Paper I, where we show that the bubbles can be formed by a recent AGN jet event which started around $1$ - $3$ Myr ago and lasted for $\sim 0.1$ - $0.5$ Myr. In the current paper, we continue to investigate the potential roles of gas viscosity and CR diffusion on the evolution of the \emph{Fermi} bubbles in the jet scenario, while directly comparing the model to {\it Fermi} observations. We refer the reader to Paper I for details of our model and assumptions. Here we simply summarize the main model assumptions, with a focus on several modifications.
      
\subsection{Equations and Assumptions}
\label{section:equation}

 \begin{figure}
\plotone{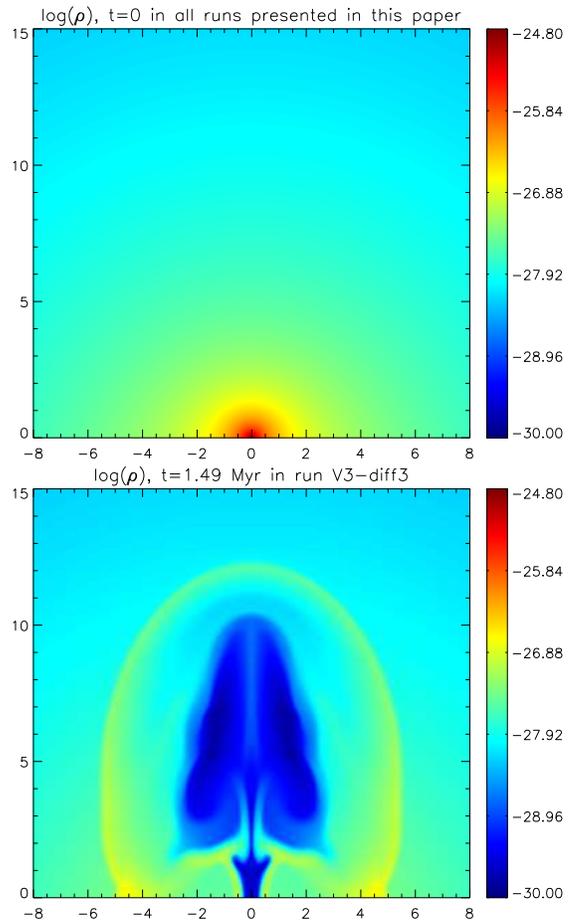}
\caption{{\it Top}: The initial density distribution of the hot thermal gas in the Galactic halo for all runs presented in this paper. Horizontal and vertical axes refer to $R$ and $z$ respectively, labeled in kpc. The hot gas is assumed to be isothermal ($T=2.4 \times 10^{6}$ K) and in hydrostatic equilibrium at time $t=0$. {\it Bottom}: The density distribution of the hot gas at $t=t_{\rm Fermi}$ in run V3-diff3 (see Table 1). The low-density cavity and the surrounding shock, both produced by the AGN jet event, are clearly seen.
}
 \label{plot1}
 \end{figure}
 
In our model, CRs and hot gas are treated as an interacting ``two-fluid" system, whose dynamical evolution may be described by the following four equations:

\begin{eqnarray}
\frac{d \rho}{d t} + \rho \nabla \cdot {\bf v} = 0,\label{hydro1}
\end{eqnarray}
\begin{eqnarray}
\rho \frac{d {\bf v}}{d t} = -\nabla (P+P_{\rm c})-\rho \nabla \Phi +\nabla \cdot {\bf \Pi},\label{hydro2}
\end{eqnarray}
\begin{eqnarray}
\frac{\partial e}{\partial t} +\nabla \cdot(e{\bf v})=-P\nabla \cdot {\bf v}+{\bf \Pi}:\nabla {\bf v}
   \rm{ ,}\label{hydro3}
   \end{eqnarray}
\begin{eqnarray}
\frac{\partial e_{\rm c}}{\partial t} +\nabla \cdot(e_{\rm c}{\bf v})=-P_{\rm c}\nabla \cdot {\bf v}+\nabla \cdot(\kappa\nabla e_{\rm c})
   \rm{ ,}\label{hydro4}   \end{eqnarray}
  \\ \nonumber
\noindent
where $d/dt \equiv \partial/\partial t+{\bf v} \cdot \nabla $ is the
Lagrangian time derivative, $\kappa$ is the CR diffusion
coefficient, {\bf $\Pi$} is the viscous stress tensor (see the Appendix), and all other variables have their usual meanings. The gas pressure $P$ and CR pressure $P_{\rm c}$ are related with the gas and CR energy density $e$ and $e_{\rm c}$ via $P=(\gamma-1)e$ and $P_{\rm c}=(\gamma_{\rm c}-1)e_{\rm c}$ respectively, where we assume $\gamma=5/3$ and $\gamma_{\rm c}=4/3$. The nature of the relativistic particles with energy density $e_{\rm c}$ is unspecified and may be electrons and/or protons with any spectra. Of course the equation of state may be somewhat harder if $e_{\rm c}$ is mainly contributed by trans-relativistic protons at $\sim 1$ GeV.
  
We assume a temporally fixed Galactic potential $\Phi$, which is contributed by three components: the bulge, disk and dark matter halo. Their properties are elaborated in Section 2.3 of Paper I. At time $t=0$ we assume that the CR energy density is zero in the Galaxy, $e_{\rm c}=0$. The hot Galactic gas is assumed to be initially isothermal with temperature $T=2.4 \times 10^{6}$ K and its density distribution is solved from the assumption of hydrostatic equilibrium. The free parameter $n_{\rm e0}$, the thermal electron number density at the origin, determines the normalization of the gas density distribution and is chosen to be $0.1$ cm$^{-3}$ throughout this paper. The resulting initial density distribution is shown in the top panel of Figure \ref{plot1}, while the current density distribution after the AGN event in a typical run V3-diff3 is shown in the bottom panel. The dependence of our results on the poorly-constrained parameter $n_{\rm e0}$ has been explicitly explored in Paper I, which shows that for higher values of $n_{\rm e0}$, more powerful jets are needed to produce the \emph{Fermi} bubbles with the same observed morphology.

We ignore radiative cooling of thermal gas, which is unimportant during the short-duration ($\lesssim 1$-$3$ Myr) of our simulations. We also neglect CR energy losses from synchrotron and IC emissions, which is a good assumption for CR protons and CR electrons at $10$ - $100$ GeV or lower energies. In this paper, we follow the evolution of the integrated CR energy density $e_{\rm c}$ (equation \ref{hydro4}), which may not be significantly affected by the CR cooling as it is probably mainly contributed by low-energy CR electrons and possibly CR protons with long lifetimes. The cooling may be important for electrons at TeV energies, and thus may significantly affect the gamma-ray spectrum if TeV electrons dominate gamma-ray emissions from the \emph{Fermi} bubbles (in particular at high latitudes). The main purpose of this paper is to reproduce the basic observed morphology of the gamma-ray bubbles, but we defer a detailed direct comparison with the data (including spectral predictions) to future work.
  
Equation \ref{hydro4} describes the evolution of CR energy density including both advection and diffusion. Note that -- as discussed in Section 2.1 of Paper I -- we ignore CR streaming, which may play a similar role as CR diffusion, transporting CRs away from local thermal plasma. Typical values of CR diffusivity $\kappa$ in the Galaxy are found to be $\kappa \sim (3-5)\times 10^{28}$ cm$^{2}$ s$^{-1}$ for CRs at about 1 GeV\citep{strong07}. However, in Paper I we show that to produce the sharp edges of the observed \emph{Fermi} bubbles, the CR diffusivity across the bubble surface must be strongly suppressed, which may occur naturally if magnetic field lines are mainly tangential on the bubble surface, as expected from magnetic draping. Previous work has explored similar effects in AGN blown bubbles in galaxy clusters, with similar results \citep{mathews07,ruszkowski08}. Thus for most runs, we choose a strongly-suppressed uniform and constant CR diffusivity $\kappa =3\times 10^{26}$ cm$^{2}$ s$^{-1}$, which ensures that CR diffusion has a negligible effect on the few-Myr evolution of our simulated bubbles. In Section 3.2, we present three additional runs where without increasing CR diffusivity outside the bubble, we increase CR diffusivity in the bubble interior to $\kappa_{\rm int} =(1-6)\times 10^{28}$ cm$^{2}$ s$^{-1}$, exploring the effect of CR diffusion on the CR distribution in the bubble interior.

Equations (\ref{hydro1}) $-$ (\ref{hydro4}) are solved in $(R, z)$ cylindrical coordinates using a 2D axisymmetric Eulerian code similar to ZEUS 2D \citep{stone92}.\footnote{Since both vorticity and magnetic flux are subject to stretching in 3D, but not in 2D, the restriction to 2D has more of an effect in MHD simulations of the KH instability, which attempt to characterize the field strength required for magnetic tension to stabilize the flow \citep{ryu00}, an issue beyond the scope of this study. Such studies generally find that a smaller initial field is required in 3D to stabilize the flow, due to greater field amplification.} In particular, we have implemented important new physics into the code, including gas viscosity (see the Appendix), CR advection and diffusion, and the dynamical interaction between hot gas and CRs. The computational grid consists of $400$ equally spaced zones in both coordinates out to $20$ kpc plus additional $100$ logarithmically-spaced zones out to $50$ kpc. The jet inflow is introduced along the $z$-axis (the rotation axis of the Galaxy) from the GC. See Paper I for the details of our jet injection method. The jet parameters are the same in all the runs presented in this paper: speed $v_{\rm jet}=3.0\times 10^{9}$ cm/s, radius $R_{\rm jet}=0.4$ kpc, duration $t_{\rm jet}=0.4$ Myr, CR energy density $e_{\rm jcr}=1.0\times 10^{-10}$ erg/cm$^{3}$, thermal gas density $\rho_{\rm j}=1.102\times10^{-28}$ g/cm$^{3}$ (density contrast $\eta=0.01$ with respect to the ambient gas), thermal energy density $e_{\rm j}=5.4\times 10^{-12}$ erg/cm$^{3}$. These jet parameters produce CR bubbles having similar morphologies as the observed \emph{Fermi} bubbles. The total power of this jet is $P_{\rm jet}\sim 8.6 \times 10^{42}$ erg s$^{-1}$, dominated by the kinetic power. The total energy injected by these two opposing jets is $2E_{\rm jet}=2P_{\rm jet}t_{\rm jet}\sim 2.17\times 10^{56}$ erg. We stop each simulation at time $t=t_{\rm Fermi}$, when the produced CR bubble reaches $z=10.5$ kpc along the jet axis. At this time, the projected CR bubble viewed in the Galactic coordinate system roughly reaches the highest latitude to which the observed \emph{Fermi} bubbles extend (see the bottom panels of Fig. \ref{plot6}). Thus $t_{\rm Fermi}$ is the predicted dynamical age of the \emph{Fermi} bubbles in each model, and depends on model parameters.

The dependence of our results on jet parameters has been explored in Paper I. In particular, we point out that the CR energy density $e_{\rm jcr}$ is roughly degenerate with thermal energy density $e_{\rm j}$ in the sense that the jet evolution depends on the total jet pressure ($e_{\rm jcr}/3+2e_{\rm j}/3$) and is not very sensitive to the specific CR pressure. Thus our current model can not uniquely predict the gamma-ray luminosity of the resulting \emph{Fermi} bubbles. However, the observed gamma ray flux may be used to constrain the particle content in the bubbles. We did some simple emission calculations in Section 3.5 of Paper I and found that (1) If the gamma-ray emission from the bubbles is mainly produced by CR electrons, the required CR electron pressure is negligible compared to the total bubble pressure, which may instead be dominated by other components, e.g., thermal gas, CR protons, or magnetic fields. (2) If the gamma-ray emission is mainly due to CR protons, the required CR proton pressure is much higher, probably dominating the total bubble pressure.

\begin{table}
 \centering
 \begin{minipage}{80mm}
  \renewcommand{\thefootnote}{\thempfootnote} 
  \caption{List of Simulations}
    \centering
  \begin{tabular}{@{}lccc}
  \hline 
  & $\mu_{\rm visc}$&{$\kappa_{\rm int}$\footnote{$\kappa_{\rm int}$ is the value of CR diffusion coefficient within the evolving CR bubble (see Sec. 3.2 for details). The CR diffusivity outside the bubble is always chosen to be $3\times10^{26}$ cm$^{2}$ s$^{-1}$, an arbitrarily-chosen small value to suppress CR diffusion across bubble edges.}}&$t_{\rm Fermi}$\\ 
  Run&(g cm$^{-1}$ s$^{-1}$)&(cm$^{2}$ s$^{-1}$)&(Myr) \\ \hline 
       V0  .................. &0&$3\times10^{26}$&1.85 \\ 
      V0.5  .............. &0.5&$3\times10^{26}$  & 1.81 \\ 
     V1  .................. & 1&$3\times10^{26}$&1.87 \\ 
       V3  .................. &3&$3\times10^{26}$&1.67 \\
     V10  ................  & 10&$3\times10^{26}$&1.60 \\ 
        V30  ................  & 30&$3\times10^{26}$&2.24 \\          
        V3-diff1  ..........  & 3&$1\times10^{28}$&1.60 \\          
       V3-diff3  ..........  &3&$3\times10^{28}$&1.49 \\ 
       V3-diff6  ..........  &3&$6\times10^{28}$&1.39 \\       
          \hline
\label{table1}
\end{tabular}
\end{minipage}
\end{table}

\subsection{The Role of Viscosity in Jet/Bubble Evolution}

  \begin{figure*}
\plotone {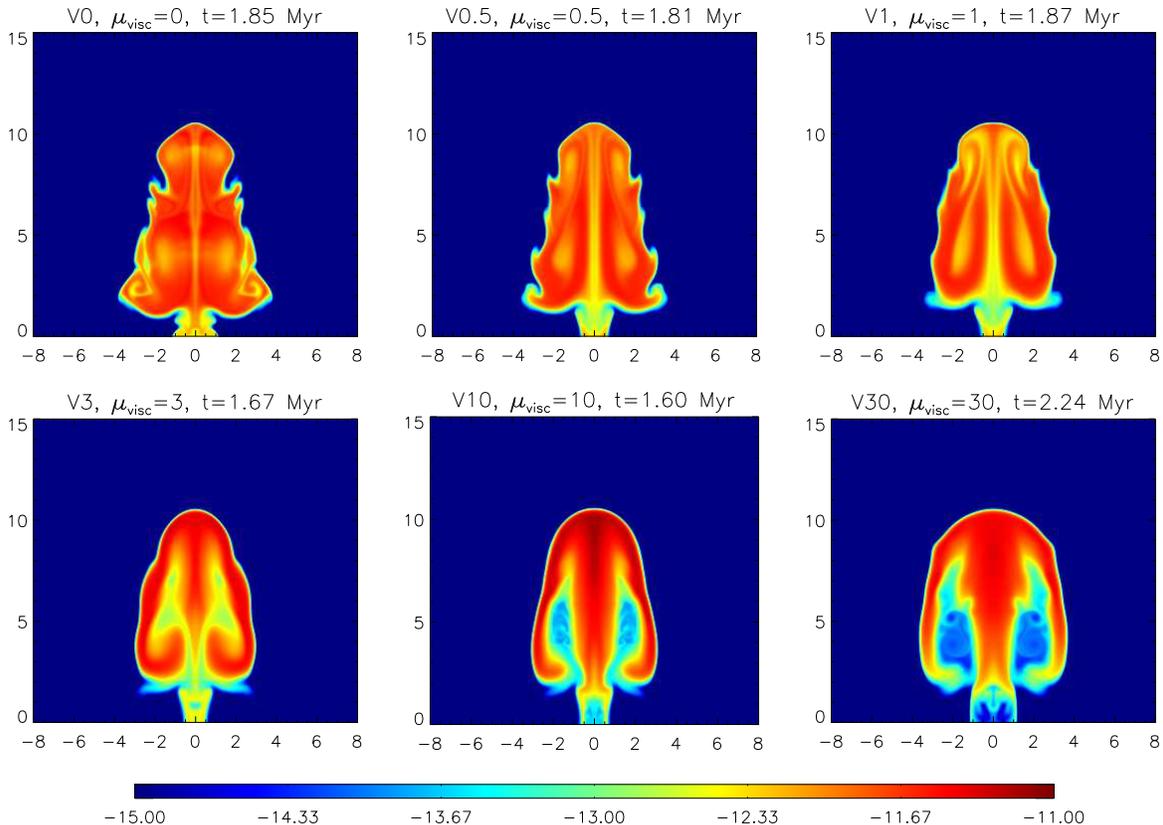}
\caption{Central slices ($16\times15$ kpc) of CR energy density in logarithmic scale in runs V0, V0d5, V1, V3, V10, and V30 at $t=t_{\rm Fermi}$, which is shown at the top of each panel for the corresponding run. Horizontal and vertical axes refer to $R$ and $z$ respectively, labeled in kpc. The stabilizing effect of viscosity on bubble edges can be clearly seen here as viscosity increases from panel to panel, and the Kelvin-Helmholtz and Rayleigh-Taylor instabilities are fully suppressed when $\mu_{\rm visc} \gtrsim 3$ g cm$^{-1}$ s$^{-1}$.}
 \label{plot2}
 \end{figure*} 
 
One of the main goals of this paper is to study the role of shear viscosity on the evolution of the \emph{Fermi} bubbles in the jet scenario. This study is mainly motivated by our previous jet simulations presented in Paper I, where the Kelvin-Helmholtz (KH) and potential Rayleigh-Taylor (RT) instabilities develop at the edges of the resulting CR bubbles, strikingly inconsistent with smooth edges of the observed \emph{Fermi} bubbles. Smooth bubble edges suggest that the instabilities are effectively suppressed by some additional physics, which works on small scales along the whole bubble surface. Viscosity is an ideal candidate mechanism to fulfill this purpose.  

In a fully ionized, unmagnetized, thermal plasma, the dynamic viscosity coefficient is (\citealt{braginskii58}; \citealt{spitzer62}):
\begin{eqnarray}
\mu_{\rm visc}=6.0\times10^{3}\left(\frac{\text{ln }\Lambda}{37}\right)^{-1}\left(\frac{T}{10^{8}\text{ K}}\right)^{5/2} \text{  g cm}^{-1}\text{ s}^{-1}\text{,} \label{equvisc}
\end{eqnarray}
where $T$ is the temperature of the plasma in Kelvin and $\text{ln }\Lambda$ is the Coulomb logarithm. One important property of the viscosity is that it increases dramatically with gas temperature ($\mu_{\rm visc} \propto T^{5/2}$). For example, the value of $\mu_{\rm visc}$ increases from $0.06$ to $6000\text{  g cm}^{-1}\text{ s}^{-1}$ when temperature increases from $10^{6}$ to $10^{8}$ K. As shown in Paper I, the AGN jet event induces a strong shock propagating into the hot halo gas, which heats the gas to tens to hundreds of keV at early times. The gas temperature drops as the gas expands into the halo, but even at the current time, the shocked gas still has temperatures of a few keV. Large viscosity in such hot gas may potentially play a significant role in the evolution of the \emph{Fermi} bubbles (more generally AGN bubbles), in particular, suppressing the development of KH and RT instabilities. Unknown magnetic fields may suppress viscosity across field lines, 
but as we show in this paper, a very small fraction (less than 1\%) of the viscosity in equation (\ref{equvisc}) is capable of suppressing these instabilities.

Unlike the theories of accretion disks, where the role of viscosity is greatly appreciated, theoretical/numerical studies of AGN jets often ignore viscosity. One reason for this neglect is the difficulty in attributing observational features directly with the effects of viscosity; the smooth edges of the \emph{Fermi} bubbles may provide an unusual opportunity to place such observational constraints. Similar smooth edges have also been observed in many radio bubbles in galaxy clusters, which motivated \citet{reynolds05} to study the role of viscosity on the evolution of buoyantly rising bubbles. These studies also found that modest levels of viscosity could stabilize Rayleigh-Taylor and Kelvin-Helmholz instabilities, allowing the bubbles to maintain their integrity. However, the numerical study of \citet{reynolds05} only considered the buoyant rise of initially static bubbles, side-stepping the initial jet-driven inflation of the bubble. As they acknowledge, besides excluding the effect of the jet on its surroundings (such as driving strong shocks), this leaves out complex internal motions within the bubble which arise during the inflation phase. By directly simulating the AGN jet, we take such effects into account. We find that internal backflows within the bubble contribute strongly to the development of fluid instabilities at the bubble surface, and cannot be ignored; indeed, viscosity plays a critical role in mitigating such backflows.

In this paper, we adopt a constant, isotropic viscosity and study how the results vary with different values of viscosity in a series of simulations (see Table 1). This simplified approach, which was also adopted by \citealt{reynolds05}, allows us to make a preliminary assessment of how bubble evolution might be affected by viscosity (see Figure \ref{plot2}). The true nature of viscosity here is highly uncertain. For instance, equation (\ref{equvisc}) describes the isotropic viscosity coefficient in an unmagnetized hot plasma, but the gas/plasma in and outside the \emph{Fermi} bubbles contains magnetic fields, which makes viscosity anisotropic, as it is enormously suppressed (by a factor $\sim 10^{23}$) across field lines. The exact value of viscosity in a turbulent medium with tangled field lines is unknown, although perhaps in analogy with thermal conduction \citep{narayan01}, values $\sim 1-30\%$ of the Braginskii-Spitzer value are plausible. The value could be considerably smaller if the field is coherent: for instance, nearly-parallel magnetic field lines at the bubble surface, as suggested by sharp bubble edges (see Sec. 3.2 of Paper I), may significantly suppress momentum transport across bubble edges, although some tangling of the field here--perhaps due to the instabilities itself--could still allow a non-negligible value. The nature and level of viscosity in the bubble interior are even more uncertain, as the thermal gas there is extremely hot and underdense; the formal Coulomb mean free path:
\begin{eqnarray}
\lambda_{\rm mfp}\sim 3 \times 10^{5} \left(\frac{T}{100\text{ keV}}\right)^{2}\left(\frac{n_{\rm e}}{10^{-5}\text{ cm}^{-3}}\right)^{-1} \text{ kpc,}
\end{eqnarray}
is so large that it is effectively collisionless.

Thus, while we compare the values of viscosity that we use to the Braginskii-Spitzer value to illustrate its magnitude, we make no claims as to its provenance. For instance, the stress could be magnetic in nature. It could also arise from anisotropic pressure, which is sensitive not just to the topology but the magnitude of magnetic fields. In a weakly collisional/collisionless plasma such as the bubble interior, pressure anisotropy $p_{\parallel} \neq p_{\perp}$ arises from conservation of the magnetic moment for each particle $\mu = m v_{\perp}^{2}/2B = {\rm const},$ which implies that any change in the field is accompanied by a change in the perpendicular pressure to keep $p_{\perp}/B \sim$const. This then triggers micro-instabilities (such as the firehose, mirror, ion cyclotron instabilities) which feed off the pressure anisotropy and pin it at marginal stability values \citep{rosin11}. The micro-instabilities change the pressure anisotropy either via an enhanced rate of collisions through an effective particle scattering mechanism, a source of effective viscosity \citep{sharma06}, or modification of the rate of strain of the magnetic field so as to cancel the pressure anisotropy created by the changing fields \citep{rosin11,schekochihin10}; the latter gives rise to a viscosity in a turbulent medium that scales as the parallel Braginskii value (and by dissipating turbulent motions, could provide significant viscous heating; \citealt{kunz11}). Viscosity in collisionless plasma may also be caused by particle scattering with magnetic irregularities and Alfven waves, which has been invoked to explain the origin of CR diffusion -- a well-known transport process in collisionless plasma. Assuming that $\mu_{\rm visc} \sim \rho \bar{v} \lambda$, the effective mean free path of proton scattering for our assumed level of viscosity is:
\begin{eqnarray*}
\lambda\sim &1& \text{ kpc} \left(\frac{\mu_{\rm visc}}{3\text{ g cm}^{-1}\text{ s}^{-1}}\right) \left(\frac{\bar{v}}{10^{8}\text{ cm s}^{-1}}\right)^{-1}  \\
& \times& \left(\frac{\rho}{10^{-29}\text{ g cm}^{-3}}\right)^{-1} \text{,} ~~~~~~~~~~~~~~~~~~~~~~~~~~~~~~~~~~ (7)
\end{eqnarray*}
where $\bar{v}$ is the kinetic velocity of protons and $\rho$ is the plasma density.

Thus, while the nature of viscosity in this context is highly uncertain, assuming an isotropic, uniform viscosity is not unreasonable. The next step would obviously be to perform MHD simulations similar to those of \citep{sharma06} for accretion disks. It would be exciting to place empirical constraints on viscosity based on comparisons of our calculations with the observed \emph{Fermi} bubbles.

In the Appendix, we explicitly present our numerical method to implement the fully compressible shear viscosity into our 2D code. The viscous runs are fairly expensive, because the time-step imposed by viscosity scales with $\rho (\Delta x)^{2}/\mu_{\rm visc}$, where $\Delta x$ is the resolution of the computational grid. In particular, the viscous time-step becomes extremely small at some small regions in the bubble interior, where the thermal gas density is very low due to the low initial jet density, the bubble expansion and viscous heating. To allow the simulations to proceed, we thus turn off viscosity in computational cells where the thermal gas density drops below $10^{-30}$ g cm$^{-3}$. This restriction only affects some small regions deep inside the bubbles, and does not appreciably affect the bubble evolution. 
 
  \begin{figure*}
\plotone {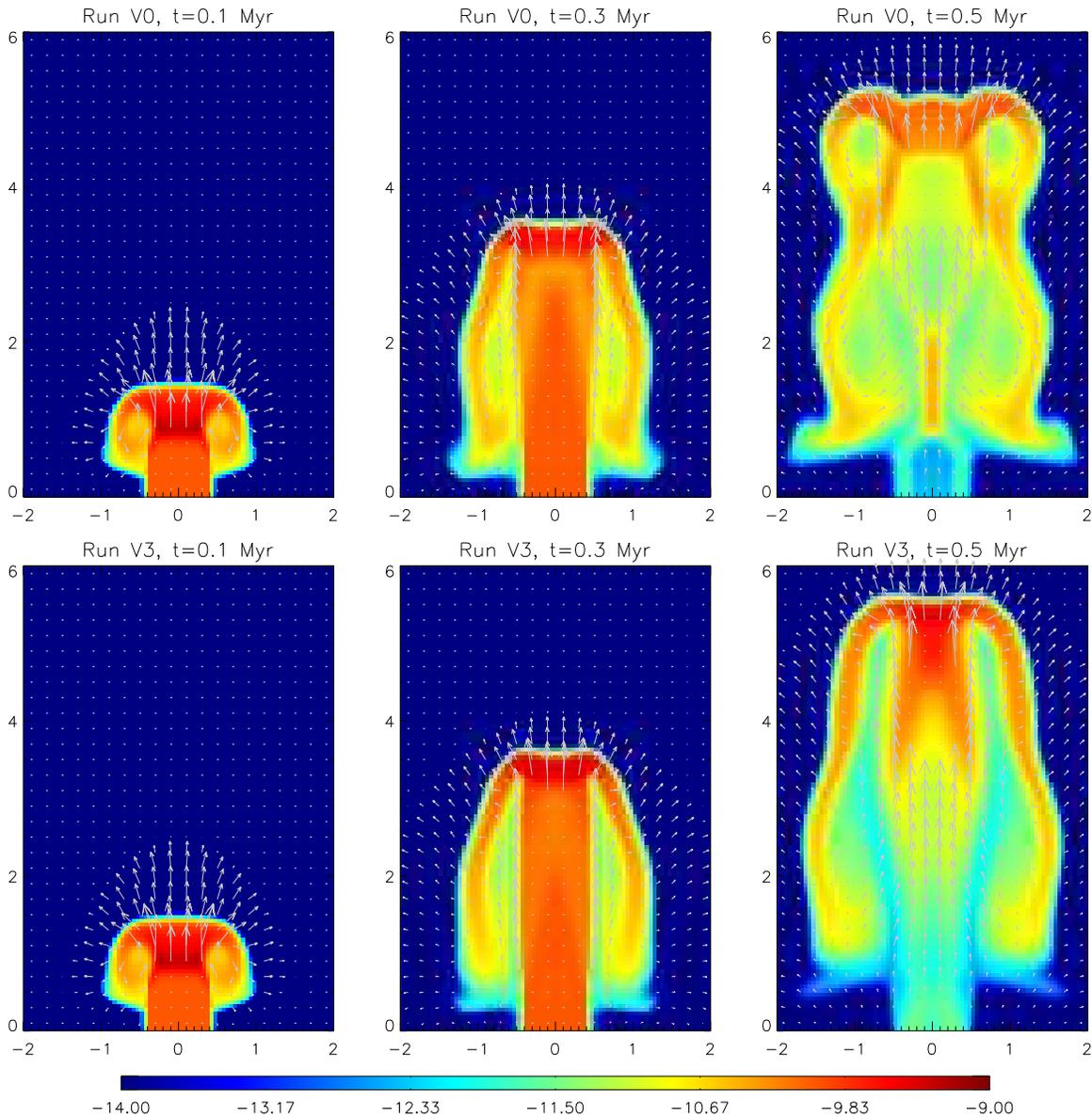} 
\caption{Central slices of CR energy density in logarithmic scale in run V0 (top panels), and V3 (bottom panels) at $t=0.1$ (left), $0.3$ (middle) and $0.5$ Myr (right). Horizontal and vertical axes refer to $R$ and $z$ respectively, labeled in kpc. Arrows superposed show gas velocity. White dots within the CR bubble indicate that the local velocity exceeds $2.5\times10^{9}$ cm/s, while those far away from the bubble indicate vanishing velocities. In the non-viscous run V0, the jet backflow at the bubble edges reaches the bottom, and then rises up in the interior, forming a large scale circulation. However, in the viscous run V3, the jet backflow stops flowing backward quickly and stays at the expanding surface without forming circulations.}
 \label{plot3}
 \end{figure*}  
 
\section{Results}
\label{section:results}

Here we present the results of our numerical calculations, which are compared directly with the gamma-ray observations of the \emph{Fermi} bubbles. Our main purpose is not to perfectly reproduce the \emph{Fermi} bubbles, but rather we aim to identity potential physical mechanisms relevant for the \emph{Fermi} bubble event. In particular, we investigate the potential roles of viscosity and CR diffusion on the bubble evolution and the CR distribution within the bubbles.

\subsection{Suppression of  KH Instabilities} 
\label{section:suppins}

To study if viscosity can indeed suppress the development of KH instabilities, we performed a series of simulations with different levels of viscosity. These runs are denoted as ``run V$\mu$'', where the viscosity coefficient for our six runs are chosen to be $\mu_{\rm visc} = 0$, 0.5, 1, 3, 10, and 30 g cm$^{-1}$ s$^{-1}$ (see Table 1). In these runs, CR diffusion plays a negligible role, as CR diffusivity is always chosen to be a small constant $\kappa=3\times10^{26}$ cm$^{2}$ s$^{-1}$ (see equation 7).

Figure \ref{plot2} shows central slices of CR energy density distribution in these runs at $t=t_{\rm Fermi}$ , when the resulting CR bubble reaches a distance of $z=10.5$ kpc along the jet direction. In the non-viscous run V0, irregularities clearly develop at the surface of the jet-induced CR bubble, indicating the significance of KH instabilities. As the viscosity coefficient increases in runs V0.5 and V1, the magnitude of surface irregularities becomes smaller. In run V3, V10, and V30, surface irregularities disappear. Thus viscosity effectively suppresses RT and KH instabilities when $\mu_{\rm visc}\gtrsim 3$ g cm$^{-1}$ s$^{-1}$. This is exactly what we expected: As the viscosity coefficient increases, the viscous stress at the bubble surface becomes more significant, effectively suppressing the growth of small perturbations, which would otherwise grow into large vortices.

As discussed in Section 2.2, the true value of viscosity in and out the \emph{Fermi} bubbles is very unclear, particularly due to the uncertain role of magnetic fields in transport processes. The values of viscosity adopted in successful runs V3, V10, and V30, respectively $\mu_{\rm visc}=3$, $10$, $30\text{  g cm}^{-1}\text{ s}^{-1}$, are much less than the Spitzer viscosity of hot gas surrounding the bubbles. This hot gas is strongly heated by the jet-induced shock, and currently has temperatures of $\sim 0.6$ -- $4\times 10^{8}$ K (even higher at earlier times). For comparison, the temperature $10^{8}$ K corresponds to the Spitzer viscosity (eq. 5) of $\sim 6000\text{  g cm}^{-1}\text{ s}^{-1}$. Thus the adopted viscosities in these runs are only around $0.1\%$ - $1\%$ of the Spitzer value in the surrounding gas. Such a low viscosity level may represent the true level of momentum transport rate across bubble edges. This is due to the suppression of transport processes across nearly-parallel magnetic field lines at the bubble surface, as discussed in Section 2.2. It is likely that near the bubble surface, the magnetic field lines are not completely tangential and a small level of field tangling allows a low level of momentum transport across the bubble surface, which is strong enough to suppress the instabilities as shown in our calculations. The importance of viscous transport inside but near the bubble surface can be seen by small values of the Reynolds number in the jet backflow:
\begin{eqnarray*}
\text{Re} \sim & 0.2&\left(\frac{\rho}{10^{-29}\text{ g cm}^{-3}}\right)\left(\frac{v}{2000\text{ km s}^{-1}}\right) \\
&\times& \left(\frac{L}{0.1 \text{ kpc}}\right)\left(\frac{\mu_{\rm visc}}{3\text{ g cm}^{-1}\text{ s}^{-1}}\right)^{-1}\text{,}
~~~~~~~~~~~ (8)
\end{eqnarray*}
where $\rho\sim 10^{-29}\text{ g cm}^{-3}$ and $L\sim 0.1$ kpc are roughly the thermal gas density and thickness of the jet backflow layer. In contrast, the Reynolds number in the ambient shock-heated gas is much larger:
\begin{eqnarray*}
\text{Re} \equiv \frac{\rho v L}{\mu_{\rm visc}}& \sim&  200\left(\frac{\rho}{10^{-28}\text{ g cm}^{-3}}\right)\left(\frac{v}{2000\text{ km/s}}\right) \\
&\times& \left(\frac{L}{10 \text{ kpc}}\right)\left(\frac{\mu_{\rm visc}}{3\text{ g cm}^{-1}\text{ s}^{-1}}\right)^{-1}\text{,}
~~~~~~ (9)
\end{eqnarray*}
where $\rho\sim 10^{-28}\text{ g cm}^{-3}$ and $L\sim 10 $ kpc are roughly the thermal gas density and velocity length scale in the shocked halo gas, indicating that viscosity at this level is dynamically unimportant there.

Viscosity also helps dissipate gas motions in the bubble interior, reducing the level of CR advection, as clearly seen in Figure \ref{plot2}. When viscosity is not important (e.g., in runs V0 and V0.5), CR advection driven by circulating gas motions transports and mixes CRs within the bubble, producing a volume-filling CR bubble. In contrast, when viscosity becomes important (in runs V3, V10 and V30), gas motions are significantly reduced, and CRs are mainly located in the jet backflow near bubble edges. In the runs with low viscosity, the jet backflow forms a large scale circulating flow in the bubble interior, as better seen in run V1 (the right-top panel), which advect CRs, producing a volume-filling CR bubble. However, in runs V3, V10, and 30, viscosity dissipates the shear motions induced by the backflow near bubble edges, preventing the formation of circulating motions in the bubble interior. 

This viscosity effect can be better seen in Figure \ref{plot3}, which compares the early-stage jet evolution in runs V0 and V3. Here the velocity field is over-plotted on the distribution of CR energy density. As clearly seen, in the non-viscous run V0, the jet backflow induces significant circulating motions, which advect CRs to the bubble interior. In contrast, in the viscous run V3, circulating motions are only seen at the very early time $t=0.1$ Myr, and at later times, the jet backflow stops flowing backward and mainly expands into the Galactic halo. One consequence of this viscosity effect is the reduction of shear motions at the bubble surface, which helps suppress the growth of KH instabilities.

 \begin{figure}
\plotone {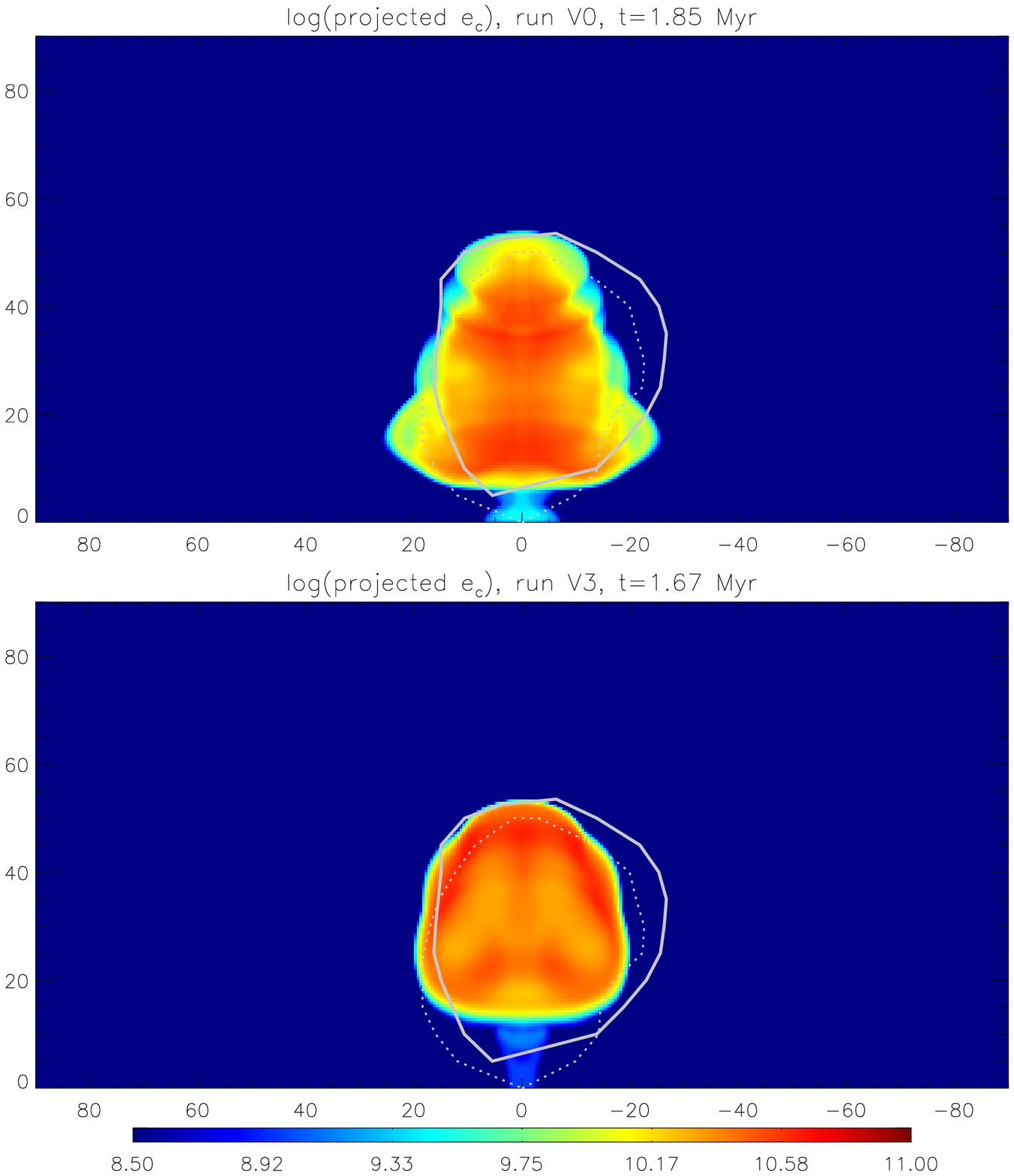} 
\caption{Line-of-sight projected CR energy density in logarithmic scale in run V0 (top) and V3 (bottom) at $t=t_{\rm Fermi}$. Horizontal and vertical axes refer to Galactic longitude and latitude respectively, labeled in degrees. The dotted region in each panel encloses the observed north \emph{Fermi} bubble, while the solid circle encloses the south \emph{Fermi} bubble. Edge irregularities are clearly seen in the non-viscous run V0, while the observed  \emph{Fermi} bubbles show smooth edges. The viscous run V3 shows smooth edges, but the gamma-ray intensity distribution is limb-brightened, inconsistent with the observed flat surface brightness.}
 \label{plot4}
 \end{figure}

\subsection{The Flat Gamma-ray Surface Brightness} 
\label{section:flatsb}

An important feature of the observed \emph{Fermi} bubbles is the approximately uniform gamma-ray surface brightness, particularly at high latitude ($|b|\gtrsim 30^{\circ}$; see \citealt{su10}). It is not trivial to form such a flat surface brightness distribution. The top panel of Figure \ref{plot4} shows the line-of-sight projected CR energy density distribution in Galactic coordinates in the non-viscous run V0 at $t=t_{\rm Fermi}$. In addition to instability-induced irregularities at bubble edges, the bubble is clearly seen to be center-brightened. Indeed, a spatially-uniform CR distribution produces a center-brightened surface brightness in projection. The roughly flat surface brightness implies that the CR density gradually increases toward the bubble edge, and particularly increases with latitude from the bubble center. 

It is of great interest to study why the CRs should concentrate near the bubble edges. CR diffusion and streaming tend to make the CR distribution more uniform, and CR advection associated with circulating motions in the non-viscous run V0 can also mix CRs deep within the bubble. \citet{mertsch11} argued that stochastic re-acceleration of CR electrons could account for the flat gamma-ray surface brightness if CR electrons are preferentially re-accelerated near bubble edges. They argued that KH instabilities seen in our non-viscous run V0 produce turbulence, resulting in CR re-acceleration preferentially near bubble edges. However, it is important to note that the instabilities seen in our non-viscous runs represent large scale turbulence while the 2nd-order Fermi acceleration mechanism is dominated by small scale turbulence.  Furthermore, generically this mechanism also produces limb brightened profiles, and the diffusion properties of the bubble interior must be adjusted to fit the data. Ultimately though, the \emph{Fermi} observations show that the bubble edges are very smooth, suggesting that the level of turbulence is low in these regions \citep{su10}.

\subsubsection{The Potential Role of Viscosity} 

\begin{figure*}
\plottwo {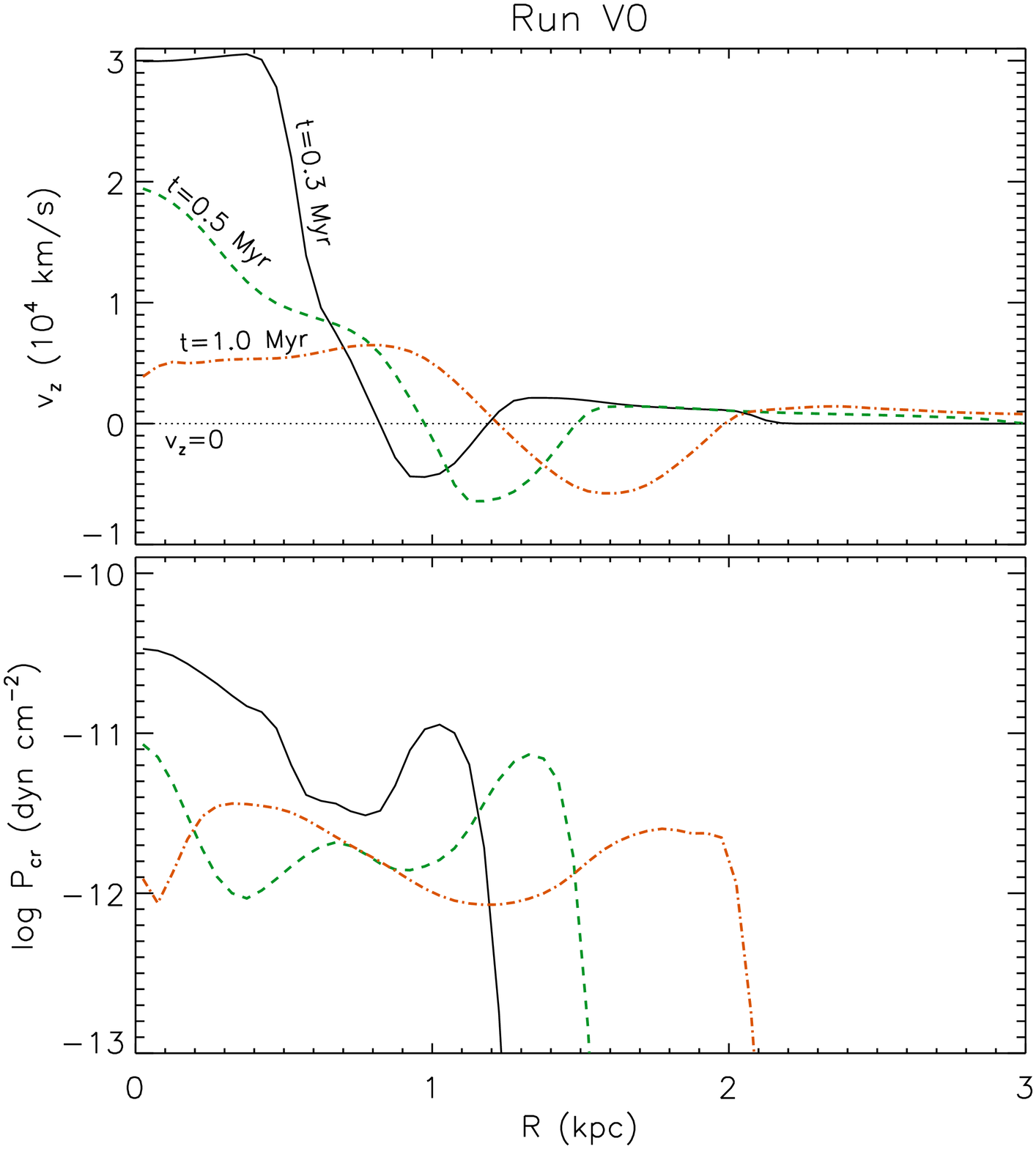}{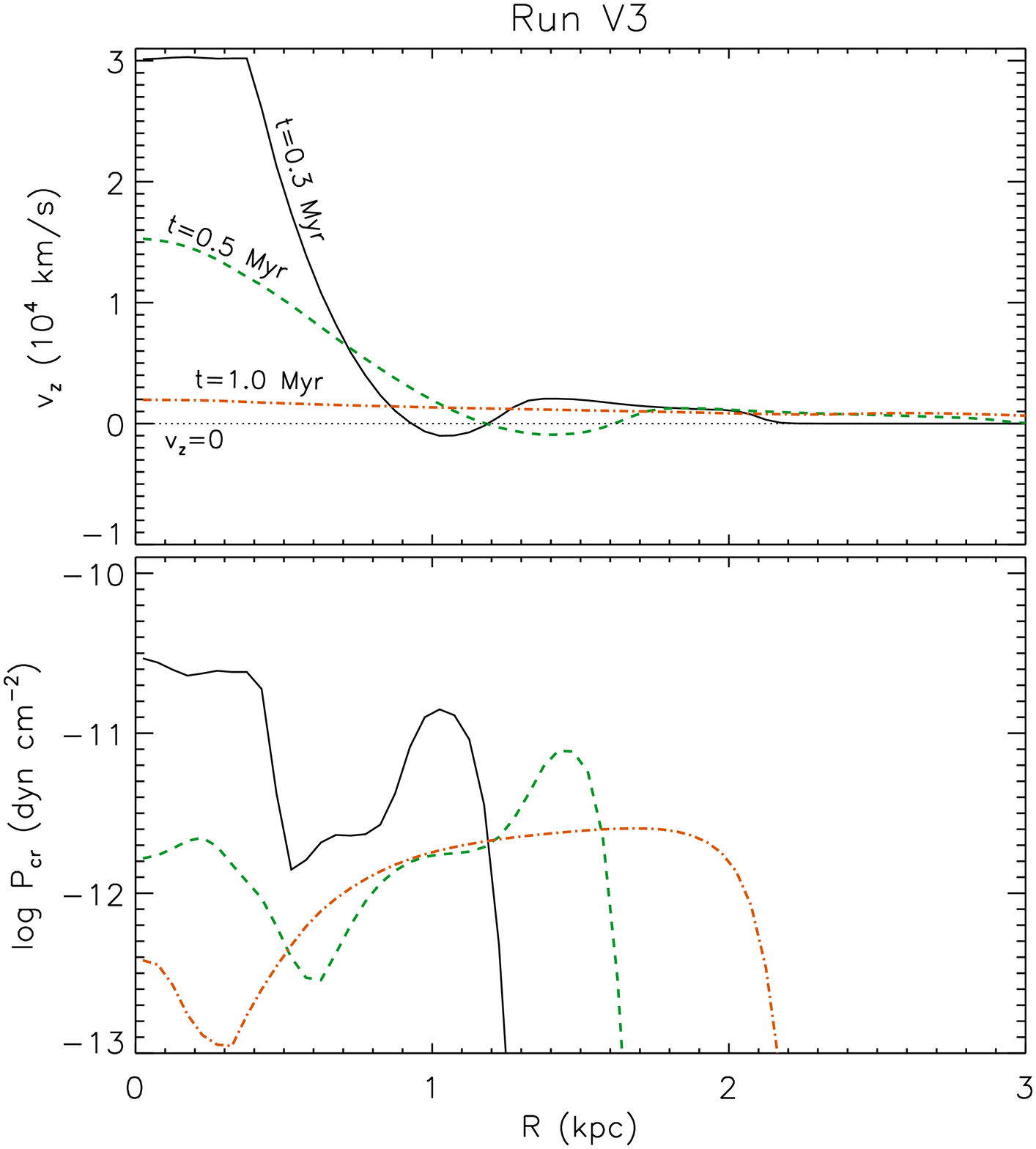} 
\caption{Variations of the $z$-component gas velocity (top) and CR pressure (bottom) along the $R$-direction (perpendicular to the jet axis) for the non-viscous run V0 (left panels) and the viscous run V3 (right panels) at $z=2$ kpc at three times $t=0.3$, $0.5$, and $1.0$ Myr. The jet backflow is represented by regions with negative values of $v_{z}$ in top panels, corresponding to CR pressure peaks at the right end of each line in the bottom panel. The backflow layer is located at the bubble surface and follows the global expansion of the CR bubble. In run V3, its backward motion is reduced by momentum kinetically transported from both the bubble interior and the ambient halo gas.}
 \label{plot5}
 \end{figure*}  
 
In the previous subsection, we show that a small level of viscosity (compared to the Spitzer viscosity in the surrounding gas) can suppress the development of KH instabilities, resulting in smooth bubble edges. Furthermore, viscosity reduces the levels of gas shear motions and the associated CR advection in the bubble interior, significantly affecting the spatial distributions of CRs and the line-of-sight projected gamma-ray intensity in the \emph{Fermi} bubbles. Due to its relatively small inertia, the jet is deflected at the top of the bubble and flows backward, transporting CRs down along the bubble boundary. In the absence of viscosity this backflow is deflected once again at the bubble bottom and returns upward in the direction of the original jet, filling most of the bubble interior with CRs. However, in viscous runs (e.g. run V3) this second upward motion is damped by viscosity and the original boundary backflow is arrested and expands with the bubble, retaining the concentration of CRs near the bubble boundary produced by the original backflow as seen in the bottom panels of Fig. \ref{plot2}. Thus, an edge-favored CR distribution, inferred from the observed flat gamma-ray intensity, is a natural consequence of shear viscosity, which is often ignored in previous jet studies. An edge-favored CR distribution may also be present in some extragalactic radio bubbles, where the observed radio synchrotron emissivity is peaked at bubble edges (\citealt{carvalho05}; \citealt{daly10}). We note that it is difficult to explain this observational feature by other physical mechanisms.

Momentum transport near the bubble surface is critical in suppressing the backward motion of the jet backflow. Figure \ref{plot5} shows variations of $v_{z}$ and CR pressure along the $R$-direction for the non-viscous run V0 (left panels) and the viscous run V3 (right panels) at $z=2$ kpc at three times $t=0.3$, $0.5$, and $1.0$ Myr. The jet backflow layer is located at the bubble surface, corresponding to the CR pressure peak at the right end of each line in bottom panels of Figure \ref{plot5}. In top panels, it is represented by regions with negative values of $v_{z}$. Due to the bubble expansion, the backflow moves to larger Galactocentric distances with time. Comparing the backflow layers in the left and right panels, it is clear that viscosity significantly suppresses the backward motion of the backflow in run V3. 

The top panels of Figure \ref{plot5} clearly show the presence of velocity gradients at both the inner and outer surfaces of the backflow, suggesting that momentum transport into the backflow from both the bubble interior and ambient gas contributes to the suppression of its backward motion in run V3, which adopts a spatially constant viscosity coefficient. It is possible that momentum transport across one surface alone may be sufficient to reduce the backflow's backward motion. In particular, strong velocity gradients are present near the inner interface of the backflow with the bubble interior as clearly seen in both the left-top (non-viscous) and right-top (viscous) panels. We speculate that momentum transport across the inner interface alone is sufficient to reduce the backflow's backward motion and thus suppress KH instabilities if momentum transport across the outer interface is fully suppressed by parallel magnetic fields. We tentatively confirm this speculation in a few additional simulations where viscosity is only allowed in the bubble interior. However, it is not easy in these simulations to accurately determine the bubble surface (i.e., to fully shut off momentum transport across the bubble surface), which is resolved by a few numerical cells. More robust conclusions can only be made by future  high resolution simulations with more advanced numerical technologies.

\subsubsection{The Potential Role of CR Diffusion} 

\begin{figure*}
\plotone {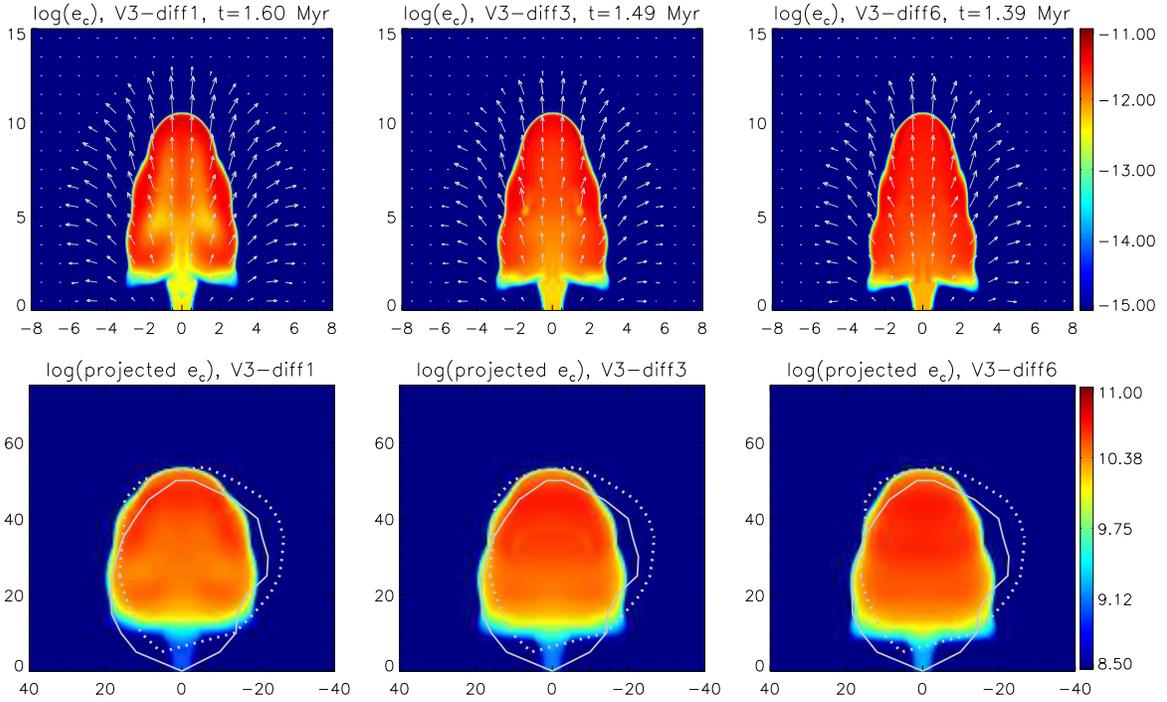} 
\caption{{Top}: central slices ($16\times15$ kpc) of log$(e_{\rm c})$ in run V3-diff1 (left), V3-diff3 (middle), and V3-diff6 (right) at $t=t_{\rm Fermi}$. Horizontal and vertical axes refer to $R$ and $z$ respectively, labeled in kpc. Arrows superposed show thermal gas velocity. {Bottom}: line-of-sight projected CR energy density in these three runs at $t=t_{\rm Fermi}$. Horizontal and vertical axes refer to Galactic longitude and latitude respectively, labeled in degree. The CR diffusivity in the bubble interior, which increases from left to right, may play an important role in explaining the flat gamma-ray surface brightness of the \emph{Fermi} bubbles.}
 \label{plot6}
 \end{figure*}  
 
The flat gamma-ray surface brightness is a very intriguing observational feature, requiring a {\it gradual} increase of CRs toward the bubble surface. As noted previously, a uniform CR distribution will give rise to a center-brightened gamma-ray surface brightness. If most CRs are concentrated at the bubble surface, the gamma-ray surface brightness will instead be limb-brightened, as clearly seen in the bottom panel of Figure \ref{plot4} which shows the distribution of line-of-sight projected CR energy density in the Galactic coordinate system in a typical viscous run V3 at $t=t_{\rm Fermi}$. We now show that including CR diffusion \emph{within} the bubbles can lead to a profile which roughly matches observations. In run V3, we choose a spatially uniform low CR diffusivity ($\kappa =3\times 10^{26}$ cm$^{2}$ s$^{-1}$), but as shown in Paper I, CR diffusion only needs be suppressed across the bubble surface to reproduce the observed sharpness of bubble edges. Such suppression could be produced by magnetic draping (particularly at early times), which produces magnetic field lines approximately tangent to the bubble surface \citep{lyutikov06, ruszkowski07, dursi08}. No such considerations apply to the bubble interior, whose field structure is likely set by the bubble inflation process and subsequent reconnection (e.g., see \citealt{braithwaite10}), probably not significantly suppressing CR diffusion there. 
We note that the details of the potential magnetic draping associated with the supersonic GC jets here may be different from those described in \citet{ruszkowski07} and \citet{dursi08}, where the draping is caused by subsonic bubble motions. Clearly, MHD jet simulations, which are beyond the scope of the current paper, would be very helpful in understanding the evolution of the magnetic field topology near the bubble surface.

To accurately study CR diffusion during the jet/bubble evolution, we need to rely upon future magnetohydrodynamic (MHD) simulations with anisotropic CR diffusion. However, within our current methodology, we can still explore the potential role of CR diffusion on the CR distribution in the bubble interior. To this end, here we present three additional runs V3-diff1, V3-diff3, and V3-diff6, in which we increase the value of CR diffusivity in the bubble interior to $\kappa_{\rm int} =1\times10^{28}$, $3\times10^{28}$ and $6\times10^{28}$ cm$^{2}$ s$^{-1}$ respectively (see Table 1 for other model parameters). We distinguish the bubble interior ($\rho < \rho_{\rm crit}$) from the outside regions ($\rho \geq \rho_{\rm crit}$) surrounding the expanding bubble by a density criterion, as the CR bubble is separated from the surrounding halo gas through a contact discontinuity, across which thermal gas density increases abruptly (see the bottom panel of Fig. \ref{plot1}). We use $\rho_{\rm crit}$ to identify gas inside the bubbles (low $\rho$, high $\kappa$) from ambient shocked gas (high $\rho$, low $\kappa$). Since in our model there are essentially no CRs outside the bubbles, the low CR diffusivity there only suppresses CR diffusion across the bubble surface. As the bubble expands quickly, the thermal gas density within the bubble drops. Therefore, the critical density ($\rho_{\rm crit}$) identifying the bubble interior should also drop with time. We choose $\rho_{\rm crit}$ to be twice the volume-averaged thermal gas density along the jet axis within the CR bubble at any time during the bubble evolution. Such a crude method produces acceptable results, as clearly seen in the top panels of Figure \ref{plot6}, which shows spatial distributions of CR energy density at $t=t_{\rm Fermi}$ in runs V3-diff1, V3-diff3, and V3-diff6. The edges of the resulting bubbles are as sharp as in the low-diffusivity run V3, and the bubble morphology is also similar in all these runs. This is consistent with what we would expect, since CR diffusion is only increased in the bubble interior while still significantly suppressed across the bubble surface. As the density jump across the bubble surface is quite large, our results are not sensitive to the specific value of $\rho_{\rm crit}$.

\begin{figure}
\plotone {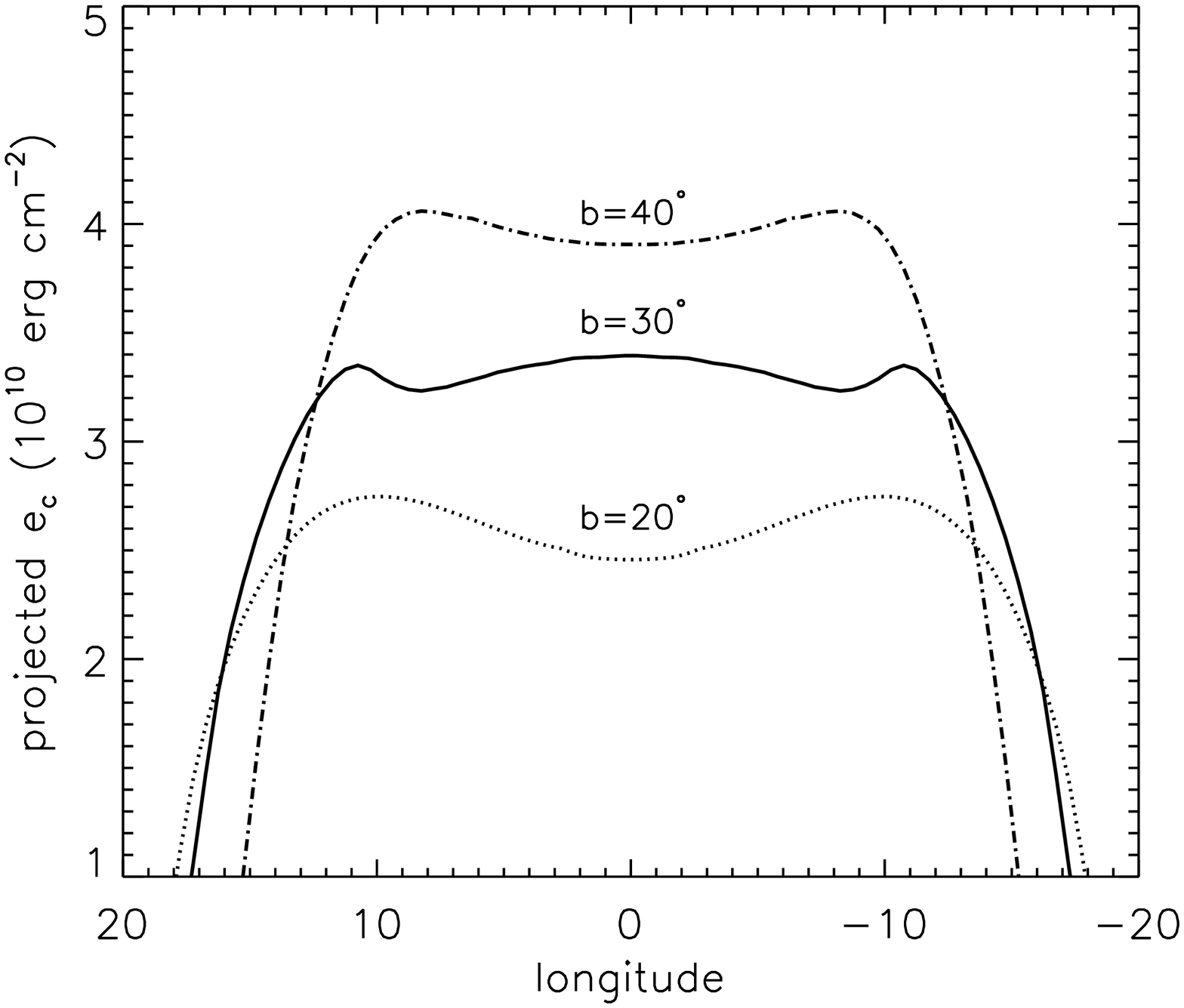} 
\caption{Line-of-sight projected CR energy density in run V3-diff3 at $t=t_{\rm Fermi}$ as a function of Galactic longitude (in degrees) at three latitudes: $b=20^{\circ}$, $30^{\circ}$, $40^{\circ}$. The projected CR distribution is roughly uniform with longitude, and slightly increases with latitude.}
 \label{plot7}
 \end{figure}

CR diffusion transports CRs near bubble edges to the bubble interior, particularly during early times when the size of the bubble is small. The typical length that CRs diffuse within a duration of $t$ is $l\sim \sqrt{\kappa t}$:
\begin{eqnarray*}
l\sim 0.3\left(\frac{\kappa}{3\times 10^{28}\text{ cm}^{2}/\text{s}}\right)^{1/2}\left(\frac{t}{1\text{ Myr}}\right)^{1/2} \text{kpc.} ~~~~(10)
\end{eqnarray*}
As $\kappa_{\rm int}$ increases, the CR distribution within the \emph{Fermi} bubbles becomes less limb-brightened and more uniform, as clearly seen in the top panels of Figure \ref{plot6} (from left to right). The bottom panels of Figure \ref{plot6} show the line-of-sight projected CR energy density in runs V3-diff1, V3-diff3, and V3-diff6 in Galactic coordinates at $t=t_{\rm Fermi}$. In run V3-diff1 with $\kappa_{\rm int} =1\times10^{28}$ cm$^{2}$ s$^{-1}$, the distribution of projected CR energy density is still limb-brightened, similar to the low-diffusivity run V3. But in runs V3-diff3 and V3-diff6 with $\kappa_{\rm int} =3\times10^{28}$ and $6\times10^{28}$ cm$^{2}$ s$^{-1}$ respectively, the projected CR energy density distribution becomes very flat at high latitude ($|b|\gtrsim 30^{\circ}$), consistent with the gamma-ray observations of the \emph{Fermi} bubbles. This  can also be seen in Figure \ref{plot7}, which shows longitudinal variations of the line-of-sight projected CR energy density in run V3-diff3 at $t=t_{\rm Fermi}$ at three latitudes: $b=20^{\circ}$, $30^{\circ}$, $40^{\circ}$. At lower latitude, the projected CR energy density is slightly lower, but higher ISRF intensities there could boost the gamma-ray IC emissivity. The flatness of the gamma ray intensity with latitude, which needs to be further corroborated by three-year or longer \emph{Fermi} observations, seems to require a fine-tuning of the latitudinal distribution of CR particles. The effects of non-uniform ISRF on the projected gamma-ray intensity will be explored in future work.
 
 \begin{figure}
\plotone {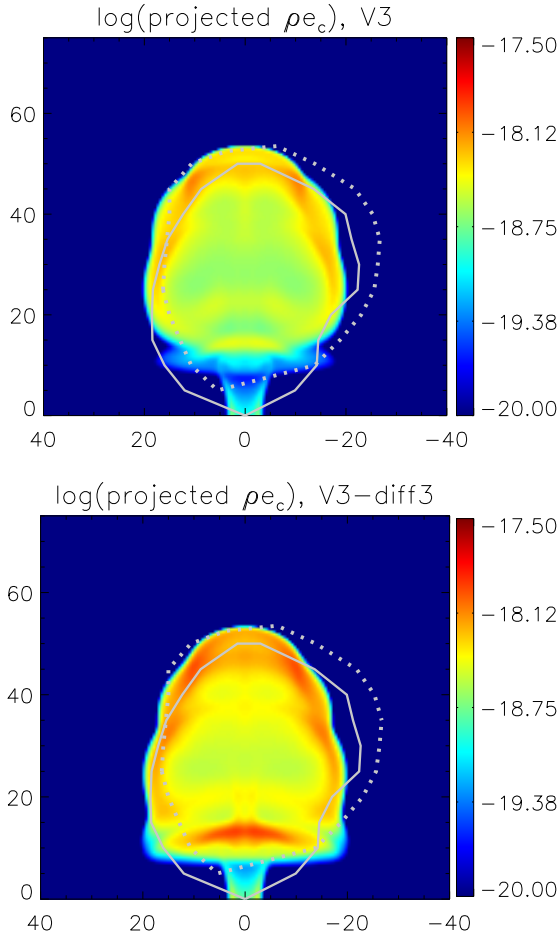} 
\caption{Line-of-sight projection of $\rho e_{\rm c}$ (a proxy for the gamma-ray surface brightness originated from CR protons) in logarithmic scale in run V3 (top panel) and  V3-diff3 (bottom) at $t=t_{\rm Fermi}$. Horizontal and vertical axes refer to Galactic longitude and latitude respectively, labeled in degrees. The gamma-ray surface brightness contributed by CR protons peaks at the jet backflow near the bubble surface, where thermal gas density is much higher than that in the bubble interior, as seen in the bottom panel of Figure \ref{plot1}.}
 \label{plot8}
 \end{figure} 
 
In the discussions above, we have mainly considered gamma ray emissions due to CR electrons through IC scattering. If CR protons are also present in the \emph{Fermi} bubbles, they will also produce gamma ray emissions. It is yet unclear if the gamma ray emission from the \emph{Fermi} bubbles is dominated by CR electrons or protons (\citealt{dobler10}; \citealt{crocker11}). The line-of-sight projections of $\rho e_{\rm c}$ at $t=t_{\rm Fermi}$ in runs V3 and V3-diff3 shown in Figure \ref{plot8} indicate that in these viscous runs (also seen in runs V3-diff1 and V3-diff6) the gamma-ray surface brightness contributed by CR protons peaks at the bubble edge, where the jet backflow is located. This edge concentration occurs because the thermal gas density in the jet backflow is much higher than that in the expanding bubble interior, as seen in the bottom panel of Figure \ref{plot1}. The gas density in the jet backflow is mainly determined by the initial jet density, which is not well constrained in our current model (see Section 3.3 and 3.4 of Paper I). But some level of fine-tuning of the initial jet density may be required if the relatively flat gamma-ray surface brightness is dominated by CR protons in our viscous jet scenario. Further studies are required to investigate if the gamma ray emission of the  \emph{Fermi} bubbles is dominated by CR electrons or protons, which is beyond the scope of the current paper.

\subsection{The Evolution of the \emph{Fermi} Bubbles} 
\label{section:evol} 

  \begin{figure*}
\plotone {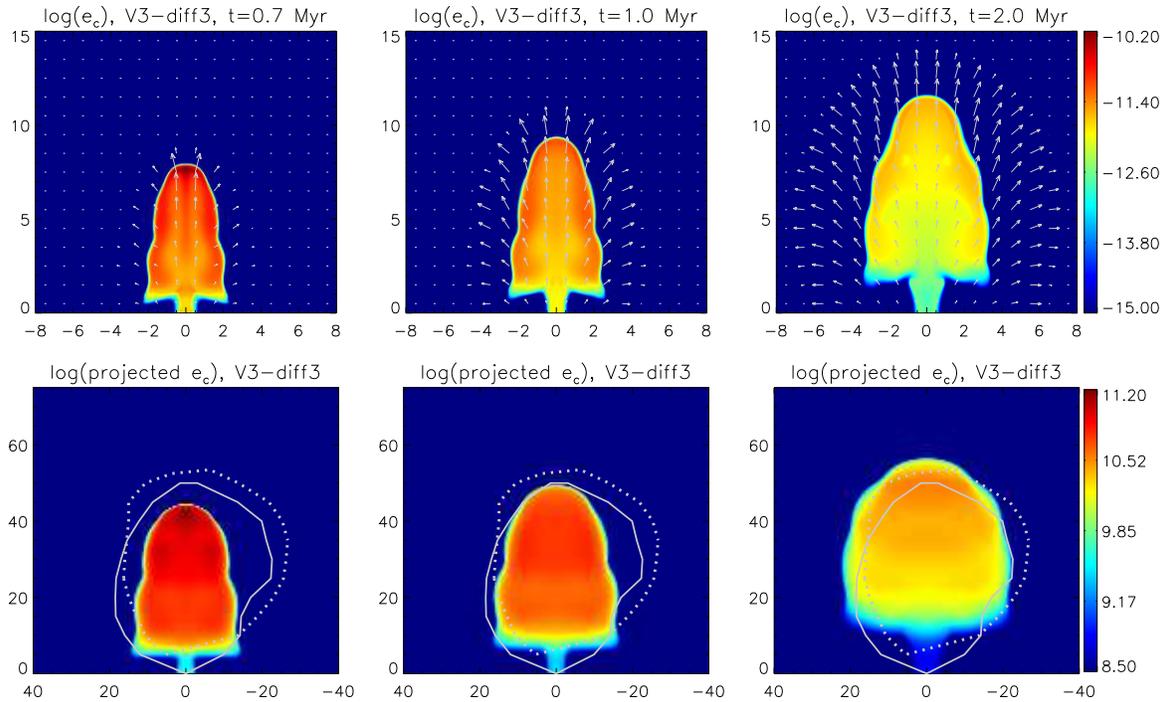} 
\caption{{\it Top}: Central slices ($16\times15$ kpc) of log$(e_{\rm c})$ in run V3-diff3 at $t=0.7$, $1$, and $2$ Myr. Horizontal and vertical axes refer to $R$ and $z$ respectively, labeled in kpc. Arrows superposed show thermal gas velocity. {\it Bottom}: The corresponding distributions of the line-of-sight projected CR energy density in run V3-diff3 at these three times in Galactic coordinates.}
 \label{plot9}
 \end{figure*}  
 
 In this subsection, we briefly comment on the temporal evolution of the \emph{Fermi} bubbles in our model. While CR diffusion transports CRs to the bubble interior, leading to a spatially-uniform CR distribution, CR advection associated with the fast bubble expansion transports CRs outward, tending to counteract this effect. Thus, a CR distribution which increases toward the bubble surface, producing a roughly flat gamma-ray surface brightness, exists for a quite long time (at least $\sim 1$ Myr) during the bubble evolution. Figure \ref{plot9} shows CR distributions (top) and the line-of-sight projected CR distributions in run V3-diff3 at times $t=0.7$, $1$, and $2$ Myr. Clearly at $t=0.7$ Myr, the CR concentration at the compressed jet tip can still be seen, but at time $t=1$ - $2$ Myr, the projected CR distribution is quite flat. Thus a flat gamma-ray surface brightness in the \emph{Fermi} bubbles in the jet scenario with viscosity is not sensitive to the exact time of observation. 

Figure \ref{plot9} also clearly shows that the CR bubble expands as it rises. The maximum expansion speed happens at the highest latitude, where it is around $2000$ km/s ($\sim 10$ degree/Myr at a distance of $\sim 8.5/\text{cos}(50^{\circ})\sim 13.2$ kpc from the solar system) at the current time $t=t_{\rm Fermi}$ ($1.49$ Myr for run V3-diff3). At lower latitudes, the bubble surface expands at slightly lower velocities ($\sim 1500$ - $2000$ km/s), which incline toward the original jet direction instead of perpendicular to the bubble surface (see velocities superposed in the top panels). Another interesting result shown in Figure \ref{plot9} is the morphological evolution of the \emph{Fermi} bubbles seen in Galactic coordinates. At early times, the bubbles are elongated in the jet (vertical) direction, but the expanding bubbles become more and more spherical and even elongated in the horizontal direction at later times. This is mainly because the top and bottom of the bubbles rise in the same direction, while the sides of the bubbles expand in opposite directions (see top panels in Fig. \ref{plot9}), and hence have larger relative velocities. The projection effect, i.e., the distance from the solar system to the bubble top is much larger than that to the bubble sides, also contributes to this morphological evolution in the Galactic coordinate system.

 \section{Summary and Implication}
\label{section:conclusion}

The \emph{Fermi Gamma-ray Space Telescope} recently detected two extended kpc-sized gamma ray bubbles in the inner Galaxy. In a companion paper (Paper I) we showed that a recent AGN jet event from the GC could produce these two bubbles with roughly the observed age, location, size and shape. However, detailed comparisons of our model with observations also reveal two discrepancies: surface irregularities and non-uniform gamma ray limb darkening. These shortcomings motivate us to further explore the jet scenario for the \emph{Fermi} bubbles in the current paper, investigating the potential roles of shear viscosity and CR diffusion on the bubble evolution.

In this paper, we performed a series of hydrodynamical jet simulations incorporating shear viscosity and CR physics (CR dynamics, advection and diffusion). In our model, the opposing jets contain two components -- thermal gas and CRs -- and naturally form two symmetric CR-filled bubbles in the inner Galactic halo. As the level of viscosity increases, it suppresses shear motions in the bubble. When the viscosity coefficient is larger than a lower limit ($\mu_{\rm visc} > \mu_{\rm min} \sim 3\text{  g cm}^{-1}\text{ s}^{-1}$ where $\mu_{\rm min}$ is even less than $1\%$ of the Spitzer viscosity of the shock-heated keV-temperature gas), shear viscosity effectively suppresses the growth of KH instabilities at the bubble surface, resulting in smooth bubble edges as observed.

The suppression of shear motions due to viscosity reduces the level of CR advection, significantly affecting the CR distribution inside the \emph{Fermi} bubbles. In non-viscous runs, the CR-carrying jet materials form an axisymmetric backflow along the bubble surface, which reaches the bubble bottom and then rises along the jet direction again, forming circulating motions in the bubble interior. The circulating motions mix CRs inside the bubble and form a CR-filled \emph{Fermi} bubble. Such a roughly-uniform CR distribution would produce a limb-darkened gamma ray intensity distribution in the line-of-sight projected Galactic coordinate system, which is inconsistent with observations. However, in viscous runs where viscosity suppresses KH instabilities, the jet backflow slows as it flows along the surface of the expanding bubbles. Thus the CR-carrying backflow mainly stays at the bubble surface, resulting in a CR distribution that increases toward the bubble edges, which is a key requirement to produce the observed flat gamma-ray surface brightness distribution. Such an edge-favored CR distribution is hard to achieve by other physical mechanisms, suggesting that viscosity may indeed play a significant role in jet evolution. 

The observed \emph{Fermi} bubbles have very sharp gamma-ray edges, indicating that CR diffusion across bubble edges is suppressed significantly below the CR diffusion rate estimated in the solar vicinity. However, if CR diffusion is also suppressed in the bubble interior, viscous runs produce limb-brightened gamma-ray bubbles in the projected Galactic coordinate system, which are also inconsistent with observations. Interestingly, if CR diffusivity in the bubble interior is close to that estimated in the solar vicinity, CR diffusion transports CRs from the bubble edges to the interior, resulting in a roughly flat projected CR distribution. Thus, by including CR diffusion within, but not across, the bubble edges, our viscous runs produce CR-filled bubbles very similar to those observed.

Given that we already invoke magnetic draping--which is inevitable and has been observed in numerous systems ranging from comets to the Sun's coronal mass ejections---to suppress CR diffusion, it might seem superfluous to invoke viscosity as a stabilizing mechanism, since magnetic tension in the drape may also stabilize the KH instability \citep{lyutikov06, ruszkowski07, dursi08}. Assuming that the results of subsonic draping described in \citet{dursi08} apply here, the field strength in the drape is independent of the ambient galactic field strength and will build up to be in rough equipartition with ram pressure (for the parameters we have chosen, it is $\sim 10$ $\mu$G); this means that the Alfven speed is of order the shear velocity, which is the requirement for the KH instability to be quenched \citep{dursi07}. We have four comments on this: (1) a further requirement for stabilization by magnetic tension is that the swept up ambient field has a coherence length $\lambda_{\rm B}$ which is larger than or of order the bubble size $R$; otherwise, the bubble will still be shredded apart, as demonstrated in MHD simulations \citep{ruszkowski07,dursi08}. While $\lambda_{\rm B}\sim R$ is plausible for bubbles in galaxy clusters, it is as yet unclear whether the halo of our Galaxy has a field coherent on $\sim 10$ kpc scales, without significant small scale irregularities. If other constraints on magnetic coherence can be placed, such as polarized emission from the drape \citep{pfrommer10}, this would indirectly probe the halo magnetic field (while there has been no detection of polarization of the
haze/bubbles by WMAP, this may be due to the significant noise in the data; \citealt{dobler12}). (2) Viscosity is also required to stabilize internal flows within the bubble, which also induce KH instabilities---indeed, our studies suggest that this source of shear is the dominant component. The influence of the swept up magnetic sheath is less clear in this case. (3)  As discussed earlier in this section, viscosity effectively suppresses the backward motions of the jet backflow and the circulating motions in the bubble interior, naturally leading to an edge-favored CR distribution, which is a key ingredient to produce the observed flat gamma ray intensity distribution. (4) The dynamics of draping in a supersonic flow where a strong shock forms - as simulated here - could be substantially different from the subsonic case simulated by previous authors.

We also studied the evolution of the \emph{Fermi} bubbles in our simulations. The top and side of the bubbles are expanding explosively, currently at speeds of around $1500-2000$ km/s, while the bottom is rising along the jet direction. In the Galactic coordinate system, the bubbles are elongated along the jet direction at early times (and the current time), and then become more spherical and even elongated in the perpendicular direction at late times. The flatness of the projected gamma ray intensity distribution may last for a quite long time ($>1$ Myr).

This work strengthens the previous result in Paper I that the \emph{Fermi} bubbles may have been recently produced by two opposing CR-carrying jets. The true level of viscosity in a magnetized low-collisional hot plasma is unclear. In particular, nearly-parallel magnetic field lines at the bubble surface may significantly suppress momentum transport across bubble edges. The level of viscosity in collisionless plasma in the bubble interior is even more uncertain. However, our calculations indicate that even a significantly-suppressed low level of viscosity ($\sim 0.1\%$ - $1\%$ of the Spitzer viscosity in the shocked surrounding gas) can play a significant role during the jet evolution, producing smooth bubble edges and an edge-favored CR distribution, which are difficult to accomplish by other mechanisms. If momentum transport across the bubble surface is fully suppressed by magnetic fields, we speculate that viscosity in the bubble interior alone may be sufficient to suppress the backward motion of the jet backflow. 


This paper suggests the potentially important roles of viscosity and CR diffusion in the Fermi bubble event; our simulations assume these to be isotropic, as is perhaps appropriate for highly tangled fields. To advance these suggestions further requires MHD simulations which are able to take into account the anisotropic nature of viscosity and CR diffusion, as well as other effects such as magnetic draping. We note that our simulations nonetheless lay the groundwork for establishing that physically reasonable values of viscosity and CR diffusion--which are very uncertain---can help explain the main features of the Fermi bubbles.

\acknowledgements

Studies of AGN feedback and the \emph{Fermi} bubbles at UC Santa Cruz are supported by NSF and NASA grants for which we are very grateful. GD has been supported by the Harvey L. Karp Discovery Award. SPO acknowledges NSF grant AST 0908480. Some discussions were initiated at a KITP workshop, supported in part by the National Science Foundation under Grant No. NSF PHY05-51164.


\begin{thebibliography}{32}
\expandafter\ifx\csname natexlab\endcsname\relax\def\natexlab#1{#1}\fi

\bibitem[{{Braginskii}(1958)}]{braginskii58}
{Braginskii}, S.~I. 1958, Soviet Journal of Experimental and Theoretical
  Physics, 6, 358

\bibitem[{{Braithwaite}(2010)}]{braithwaite10}
{Braithwaite}, J. 2010, \mnras, 406, 705

\bibitem[{Carvalho} {et~al.}(2005)]{carvalho05}
{Carvalho}, J. C., {Daly}, R. A., {Mory}, M. P., \& {O'Dea}, C. P. 2005, 
\apj, 620, 126

\bibitem[{{Cheng} {et~al.}(2011){Cheng}, {Chernyshov}, {Dogiel}, {Ko}, \&
  {Ip}}]{cheng11}
{Cheng}, K.-S., {Chernyshov}, D.~O., {Dogiel}, V.~A., {Ko}, C.-M., \& {Ip},
  W.-H. 2011, \apjl, 731, L17+

\bibitem[{{Crocker} \& {Aharonian}(2011)}]{crocker11}
{Crocker}, R.~M., \& {Aharonian}, F. 2011, Physical Review Letters, 106, 101102

\bibitem[{{Daly} {et~al.}(2010){Daly}, {Kharb}, {O'Dea}, {Baum}, {Mory},
  {McKane}, {Altenderfer}, \& {Beury}}]{daly10}
{Daly}, R.~A., {Kharb}, P., {O'Dea}, C.~P., {Baum}, S.~A., {Mory}, M.~P.,
  {McKane}, J., {Altenderfer}, C., \& {Beury}, M. 2010, \apjs, 187, 1
  
\bibitem[{{Dobler}(2012)}]{dobler12}
{Dobler}, G. 2012, ApJ, 750, 17

\bibitem[{{Dobler} {et~al.}(2011){Dobler}, {Cholis}, \& {Weiner}}]{dobler11}
{Dobler}, G., {Cholis}, I., \& {Weiner}, N. 2011, ApJ, 741, 25

\bibitem[{{Dobler} \& {Finkbeiner}(2008)}]{dobler08}
{Dobler}, G., \& {Finkbeiner}, D.~P. 2008, \apj, 680, 1222

\bibitem[{{Dobler} {et~al.}(2010){Dobler}, {Finkbeiner}, {Cholis}, {Slatyer},
  \& {Weiner}}]{dobler10}
{Dobler}, G., {Finkbeiner}, D.~P., {Cholis}, I., {Slatyer}, T., \& {Weiner}, N.
  2010, \apj, 717, 825

\bibitem[{{Dursi}(2007)}]{dursi07}
{Dursi}, L.~J. 2007, \apj, 670, 221

\bibitem[{{Dursi} \& {Pfrommer}(2008)}]{dursi08}
{Dursi}, L.~J., \& {Pfrommer}, C. 2008, \apj, 677, 993

\bibitem[{{Finkbeiner}(2004)}]{finkbeiner04a}
{Finkbeiner}, D.~P. 2004, \apj, 614, 186

\bibitem[{{Guo} \& {Mathews}(2012)}]{guo12}
{Guo}, F., \& {Mathews}, W.~G. 2012, ApJ in press (arXiv: 1103.0055)

\bibitem[{{Kaiser} {et~al.}(2005){Kaiser}, {Pavlovski}, {Pope}, \&
  {Fangohr}}]{kaiser05}
{Kaiser}, C.~R., {Pavlovski}, G., {Pope}, E.~C.~D., \& {Fangohr}, H. 2005,
  \mnras, 359, 493

\bibitem[{{Kunz} {et~al.}(2011){Kunz}, {Schekochihin}, {Cowley}, {Binney}, \&
  {Sanders}}]{kunz11}
{Kunz}, M.~W., {Schekochihin}, A.~A., {Cowley}, S.~C., {Binney}, J.~J., \&
  {Sanders}, J.~S. 2011, \mnras, 410, 2446

\bibitem[{{Lyutikov}(2006)}]{lyutikov06}
{Lyutikov}, M. 2006, \mnras, 373, 73

\bibitem[{{Mathews} \& {Brighenti}(2007)}]{mathews07}
{Mathews}, W.~G., \& {Brighenti}, F. 2007, \apj, 660, 1137

\bibitem[{{McNamara} \& {Nulsen}(2007)}]{mcnamara07}
{McNamara}, B.~R., \& {Nulsen}, P.~E.~J. 2007, \araa, 45, 117

\bibitem[{{Mertsch} \& {Sarkar}(2011)}]{mertsch11}
{Mertsch}, P., \& {Sarkar}, S. 2011, Phys. Rev. Lett., 107, 091101

\bibitem[{{Narayan} \& {Medvedev}(2001)}]{narayan01}
{Narayan}, R., \& {Medvedev}, M.~V. 2001, \apjl, 562, L129

\bibitem[{{Pfrommer} \& {Jonathan Dursi}(2010)}]{pfrommer10}
{Pfrommer}, C., \& {Jonathan Dursi}, L. 2010, Nature Physics, 6, 520

\bibitem[{{Reynolds} {et~al.}(2005){Reynolds}, {McKernan}, {Fabian}, {Stone},
  \& {Vernaleo}}]{reynolds05}
{Reynolds}, C.~S., {McKernan}, B., {Fabian}, A.~C., {Stone}, J.~M., \&
  {Vernaleo}, J.~C. 2005, \mnras, 357, 242

\bibitem[{{Rosin} {et~al.}(2011){Rosin}, {Schekochihin}, {Rincon}, \&
  {Cowley}}]{rosin11}
{Rosin}, M.~S., {Schekochihin}, A.~A., {Rincon}, F., \& {Cowley}, S.~C. 2011,
  \mnras, 413, 7

\bibitem[{{Ruszkowski} {et~al.}(2008){Ruszkowski}, {En{\ss}lin}, {Br{\"u}ggen},
  {Begelman}, \& {Churazov}}]{ruszkowski08}
{Ruszkowski}, M., {En{\ss}lin}, T.~A., {Br{\"u}ggen}, M., {Begelman}, M.~C., \&
  {Churazov}, E. 2008, \mnras, 383, 1359

\bibitem[{{Ruszkowski} {et~al.}(2007){Ruszkowski}, {En{\ss}lin}, {Br{\"u}ggen},
  {Heinz}, \& {Pfrommer}}]{ruszkowski07}
{Ruszkowski}, M., {En{\ss}lin}, T.~A., {Br{\"u}ggen}, M., {Heinz}, S., \&
  {Pfrommer}, C. 2007, \mnras, 378, 662

\bibitem[{{Ryu} {et~al.}(2000){Ryu}, {Jones}, \& {Frank}}]{ryu00}
{Ryu}, D., {Jones}, T.~W., \& {Frank}, A. 2000, \apj, 545, 475

\bibitem[{{Schekochihin} {et~al.}(2010){Schekochihin}, {Cowley}, {Rincon}, \&
  {Rosin}}]{schekochihin10}
{Schekochihin}, A.~A., {Cowley}, S.~C., {Rincon}, F., \& {Rosin}, M.~S. 2010,
  \mnras, 405, 291

\bibitem[{{Sharma} {et~al.}(2006){Sharma}, {Hammett}, {Quataert}, \&
  {Stone}}]{sharma06}
{Sharma}, P., {Hammett}, G.~W., {Quataert}, E., \& {Stone}, J.~M. 2006, \apj,
  637, 952

\bibitem[{{Spitzer}(1962)}]{spitzer62}
{Spitzer}, L. 1962, {Physics of Fully Ionized Gases, 2nd edition,}
  (Interscience, New York)

\bibitem[{{Stone} \& {Norman}(1992)}]{stone92}
{Stone}, J.~M., \& {Norman}, M.~L. 1992, \apjs, 80, 753

\bibitem[{{Strong} {et~al.}(2007){Strong}, {Moskalenko}, \&
  {Ptuskin}}]{strong07}
{Strong}, A.~W., {Moskalenko}, I.~V., \& {Ptuskin}, V.~S. 2007, Annual Review
  of Nuclear and Particle Science, 57, 285

\bibitem[{{Su} {et~al.}(2010){Su}, {Slatyer}, \& {Finkbeiner}}]{su10}
{Su}, M., {Slatyer}, T.~R., \& {Finkbeiner}, D.~P. 2010, \apj, 724, 1044

\bibitem[{{Zubovas} {et~al.}(2011){Zubovas}, {King}, \&
  {Nayakshin}}]{zubovas11}
{Zubovas}, K., {King}, A.~R., \& {Nayakshin}, S. 2011, \mnras, 415, L21

\end{thebibliography}

\appendix
\section{Implementation of Compressible Viscosity}

In this Appendix, we explicitly present the compressible viscous terms in CR-hydro equations \ref{hydro2} and \ref{hydro3} and briefly show how they are implemented in our code. We only consider shear viscosity and neglect bulk viscosity. In Cartesian coordinate systems, the viscous stress tensor ${\bf \Pi}$ can be written as
\begin{eqnarray}
\Pi_{\rm ij}=\mu_{\rm visc}\left(\frac{\partial v_{\rm i}}{\partial x_{\rm j}}+\frac{\partial v_{\rm j}}{\partial x_{\rm i}}-\frac{2}{3}\delta_{\rm ij}\nabla \cdot {\bf v}\right)
 {\rm .}
   \end{eqnarray}
However, in the cylindrical coordinate system $(R, \theta, z)$, the expression for ${\bf \Pi}$ is much more complicated. In this paper, we assume axisymmetry:
\begin{eqnarray}
\frac{\partial}{\partial \theta}=0  \text{~~~~~and~~~~~}v_{\theta}=0 \text{,}
   \end{eqnarray}
which significantly simplifies the formulae for the components of ${\bf \Pi}$. 

We first calculate the tensor of the velocity gradient $\nabla {\bf v}$ in physical units:
\begin{eqnarray}
\nabla {\bf v}=\left(  
{\begin{array}{ccc}
\partial v_{\rm R}/\partial R & ~~~~0~~~~&~~~~\partial v_{\rm R}/\partial z\\
0 & ~~~~v_{\rm R}/R~~~~ &~~~~0\\
\partial v_{\rm z}/\partial R &~~~~ 0~~~~ &~~~~\partial v_{\rm z}/\partial z\\
\end{array}}
    \right)
 \end{eqnarray}
and define the tensor ${\bf A}$: ${\bf A}=[\nabla {\bf v}+(\nabla {\bf v})^{\rm tr}]/2$, where $(\nabla {\bf v})^{\rm tr}$ is the transpose of $\nabla {\bf v}$. Then the viscous stress tensor can be written as (in physical units):
\begin{eqnarray}
{\bf \Pi}=2\mu_{\rm visc}[{\bf A}-\frac{1}{3}(\nabla \cdot {\bf v}){\bf I}]
 \text{,}
 \end{eqnarray}
where ${\bf I}$ is the identity matrix $I_{\rm ij}=\delta_{\rm ij}$, and the divergence of gas velocity is
\begin{eqnarray}
\nabla \cdot {\bf v}=\frac{1}{R}\frac{\partial}{\partial R}(Rv_{\rm R})+\frac{\partial v_{\rm z}}{\partial z}
 \text{.}
 \end{eqnarray}
After calculating all non-zero components of ${\bf \Pi}$ for each simulation zone, we finally calculate the viscous term $\nabla \cdot {\bf \Pi}$ in the gas momentum equation \ref{hydro2} and ${\bf \Pi}:\nabla {\bf v}$ in the gas energy equation \ref{hydro3}:
\begin{eqnarray}
\nabla \cdot {\bf \Pi}=\left[ \frac{\partial \Pi_{\rm RR}}{\partial R}+\frac{1}{R}(\Pi_{\rm RR}-\Pi_{\theta \theta})+\frac{\partial \Pi_{\rm zR}}{\partial z}\text{, }0\text{, }\frac{\partial \Pi_{\rm Rz}}{\partial R}+\frac{\Pi_{\rm Rz}}{R}+\frac{\partial \Pi_{\rm zz}}{\partial z}\right]
 \text{,}
  \end{eqnarray}
 \begin{eqnarray} 
 {\bf \Pi}:\nabla {\bf v}={\bf \Pi}:{\bf A}=\sum_{\rm i, j} \Pi_{\rm ij}A_{\rm ij}
 \end{eqnarray}
 Since viscosity is explicitly implemented in our code, the time-step is constrained to ensure numerical stability:
  \begin{eqnarray}
dt\leq dt_{\rm visc}=\mathcal{C}_{\rm visc}\text{ min }\left[\rho_{\rm i,j}\frac{dR_{\rm i}^{2}}{\mu_{\rm visc}},\rho_{\rm i,j}\frac{dz_{\rm j}^{2}}{\mu_{\rm visc}}\right]
 \text{,}
 \end{eqnarray} 
where $\rho_{\rm i,j}$ is the gas density in the grid cell $(R_{\rm i},z_{\rm j})$, the constant $\mathcal{C}_{\rm visc}$ is chosen to be $\mathcal{C}_{\rm visc}=0.3$, and the minimization occurs over all grid zones.   

\label{lastpage}

\end{document}